\documentclass{article}
\usepackage{amssymb}
\usepackage{amsfonts}
\usepackage{amsmath}

 \evensidemargin0in \oddsidemargin0in \topmargin10pt
\begin{document}

\title{New application of Dirac's representation: N-mode squeezing enhanced
operator and squeezed state \thanks{{\small Work was supported by the
National Natural Science Foundation of China under grants 10775097 and
10874174.}}}
\author{Xue-xiang Xu$^{1}$, Li-yun Hu$^{1,2}$\thanks{{\small Corresponding
author. E-mail addresses: hlyun2008@126.com.}} and Hong-yi Fan$^{1}$ \\
$^{1}${\small Department of Physics, Shanghai Jiao Tong University,
Shanghai, 200240, China}\\
$^{2}${\small College of Physics \& Communication Electronics, Jiangxi
Normal University, Nanchang 330022, China}}
\maketitle

\begin{abstract}
{\small It is known that }$\exp \left[ \mathtt{i}\lambda \left( Q_{1}P_{1}-%
\mathtt{i}/2\right) \right] ${\small \ is a unitary single-mode squeezing
operator, where }$Q_{1}${\small ,}$P_{1}${\small \ are the coordinate and
momentum operators, respectively. In this paper we employ Dirac's coordinate
representation to prove that the exponential operator }$S_{n}\equiv \exp [%
\mathtt{i}\lambda \sum \limits_{i=1}^{n}(Q_{i}P_{i+1}+Q_{i+1}P_{i}))],$%
{\small \ (}$Q_{n+1}=Q_{1}${\small ,\ }$P_{n+1}=P_{1}${\small ),\ is a
n-mode squeezing operator which enhances the standard squeezing. By virtue
of the technique of integration within an ordered product of operators we
derive }$S_{n}${\small 's normally ordered expansion and obtain new n-mode
squeezed vacuum states, its Wigner function is calculated by using the Weyl
ordering invariance under similar transformations.}
\end{abstract}

\section{Introduction}

Squeezed state has been a hot topic in quantum optics since Stoler \cite{1}
put forward the concept of the optical squeezing in 1970's. $S_{1}=$ $\exp %
\left[ \mathtt{i}\lambda \left( Q_{1}P_{1}-\mathtt{i}/2\right) \right] $ is
a unitary single-mode squeezing operator, where $Q_{1}$, $P_{1}$ are the
coordinate and momentum operators, respectively, $\lambda $ is a squeezing
parameter. Their variances in the squeezed state $S_{1}\left \vert
0\right \rangle =$sech$^{1/2}\lambda \exp \left[ -\frac{1}{2}a_{1}^{\dagger
2}\tanh \lambda \right] \left \vert 0\right \rangle $ are
\begin{equation*}
\Delta Q_{1}=\frac{1}{4}e^{2\lambda },\text{ }\Delta P_{1}=\frac{1}{4}%
e^{-2\lambda },\text{ }(\Delta Q_{1})(\Delta P_{1})=\frac{1}{4}.
\end{equation*}%
Some generalized squeezed state have been proposed since then. Among them
the two-mode squeezed state not only exhibits squeezing, but also quantum
entanglement between the idle-mode and the signal-mode in frequency domain,
therefore is a typical entangled states of continuous variable. In recent
years, various entangled states have attracted considerable attention and
interests of physists because of their potential uses in quantum
communication \cite{2}. Theoretically, the two-mode squeezed state is
constructed by acting the two-mode squeezing operator $S_{2}=\exp [\lambda
(a_{1}a_{2}-a_{1}^{\dagger }a_{2}^{\dagger })]$ on the two-mode vacuum state
$\left \vert 00\right \rangle $\cite{3,4,5},%
\begin{equation}
S_{2}\left \vert 00\right \rangle =\text{sech}\lambda \exp \left[
-a_{1}^{\dagger }a_{2}^{\dagger }\tanh \lambda \right] \left \vert
00\right \rangle .  \label{1}
\end{equation}%
We also have $S_{2}=\exp \left[ \mathtt{i}\lambda \left(
Q_{1}P_{2}+Q_{2}P_{1}\right) \right] ,$ where $Q_{i}$ and $P_{i}$ are the
coordinate and momentum operators related to Bose operators ($%
a_{i},a_{i}^{\dagger }$) by

\begin{equation}
Q_{i}=(a_{i}+a_{i}^{\dagger })/\sqrt{2},\ P_{i}=(a_{i}-a_{i}^{\dagger })/(%
\sqrt{2}\mathtt{i})  \label{4}
\end{equation}
In the state $S_{2}\left \vert 00\right \rangle $, the variances of the
two-mode quadrature operators of light field,

\begin{equation}
\mathfrak{X}=(Q_{1}+Q_{2})/2,\text{ }\mathfrak{P}=(P_{1}+P_{2})/2,\text{ \ }[%
\mathfrak{X},\mathfrak{P}]=\frac{\mathtt{i}}{2},  \label{2}
\end{equation}%
take the standard form, i.e.,

\begin{equation}
\left \langle 00\right \vert S_{2}^{\dagger }\mathfrak{X}^{2}S_{2}\left
\vert 00\right \rangle =\frac{1}{4}e^{-2\lambda },\text{ \ }\left \langle
00\right \vert S_{2}^{\dagger }\mathfrak{P}^{2}S_{2}\left \vert 00\right
\rangle =\frac{1}{4}e^{2\lambda },\text{ \ and }(\Delta \mathfrak{X})(\Delta
\mathfrak{P})=\frac{1}{4}.  \label{3}
\end{equation}%
On the other hand, the two-mode squeezing operator has a neat and natural
representation in the entangled state $\left \vert \eta \right \rangle $
representation \cite{6},%
\begin{equation}
S_{2}=\int \frac{d^{2}\eta }{\pi \mu }\left \vert \frac{\eta }{\mu }\right
\rangle \left \langle \eta \right \vert ,
\end{equation}%
where%
\begin{equation}
\left \vert \eta \right \rangle =\exp (-\frac{1}{2}\left \vert \eta \right
\vert ^{2}+\eta a_{1}^{\dagger }-\eta ^{\ast }a_{2}^{\dagger
}+a_{1}^{\dagger }a_{2}^{\dagger })\left \vert 00\right \rangle ,
\end{equation}%
makes up a complete set%
\begin{equation*}
\int \frac{d^{2}\eta }{\pi }\left \vert \eta \right \rangle \left \langle
\eta \right \vert =1.
\end{equation*}%
$\left \vert \eta \right \rangle $ was constructed according to the idea of
quantum entanglement innitiated by Einstein, Podolsky and Rosen in their
argument that quantum mechanics is incomplete \cite{7}.

An interesting question naturally arises: is the $n$-mode exponential
operator

\begin{equation}
S_{n}\equiv \exp \left[ \mathtt{i}\lambda
\sum_{i=1}^{n}(Q_{i}P_{i+1}+Q_{i+1}P_{i})\right] ,\text{ \ }(Q_{n+1}=Q_{1},\
P_{n+1}=P_{1}),\ n\geqslant 2,  \label{6}
\end{equation}%
a squeezing operator? If yes, what kind of squeezing for $n$-mode
quadratures of field it can engenders? To answer these questions we must
know what is the normally ordered expansion of $S_{n}$ and what is the state
$S_{n}\left \vert \mathbf{0}\right \rangle $ ($\left \vert \mathbf{0}%
\right \rangle $ is the n-mode vacuum state)? In this work we shall analyse $%
S_{n}$ in detail. But how to disentangle the exponential of $S_{n}?$ Since
the terms in the set $Q_{i}P_{i+1}\ $and $Q_{i+1}P_{i}$ ($i=1,2,\cdots ,n$)
do not make up a closed Lie algebra, the problem of what is $S_{n}$'s
normally ordered form seems difficult. Thus we appeal to Dirac's coordinate
representation and the technique of integration within an ordered product
(IWOP) of operators \cite{8,8a} to solve this problem. Our work is arranged
as follows: firstly we use the IWOP technique to derive the normally ordered
expansion of $S_{n}$ and obtain the explicit form of$\ S_{n}\left \vert
\mathbf{0}\right \rangle $; then we examine the variances of the $n$-mode
quadrature operators in the state $S_{n}\left \vert \mathbf{0}\right \rangle $%
, we find that $S_{n}$ causes squeezing which is stronger than the standard
squeezing. Thus $S_{n}$ is an $n$-mode squeezing-enhanced operator. The
Wigner function of $S_{n}\left \vert \mathbf{0}\right \rangle $ is calculated
by using the Weyl ordering invariance under similar transformations. Some
examples are discussed in the last section.

\section{Normal Product Form of $S_{n}$ derived by Dirac's coordinate
representation }

In order to disentangle operator $S_{n}$, let $A$ be%
\begin{equation}
A=\left(
\begin{array}{ccccc}
0 & 1 & 0 & \cdots  & 1 \\
1 & 0 & 1 & \cdots  & 0 \\
0 & 1 & 0 & \ddots  & 0 \\
\vdots  & \vdots  & \ddots  & \ddots  & \vdots  \\
1 & 0 & \cdots  & 1 & 0%
\end{array}%
\right) ,  \label{7}
\end{equation}%
then $S_{n}$ in (\ref{6}) is compactly expressed as%
\begin{equation}
S_{n}=\exp [\mathtt{i}\lambda Q_{i}A_{ij}P_{j}],  \label{8}
\end{equation}%
here and henceforth the repeated indices represent Einstein's summation
notation. Using the Baker-Hausdorff formula,%
\begin{equation*}
e^{A}Be^{-A}=B+\left[ A,B\right] +\frac{1}{2!}\left[ A,\left[ A,B\right] %
\right] +\frac{1}{3!}\left[ A,\left[ A,\left[ A,B\right] \right] \right]
+\cdots ,
\end{equation*}%
we have
\begin{eqnarray}
S_{n}^{-1}Q_{k}S_{n} &=&Q_{k}-\lambda Q_{i}A_{ik}+\frac{1}{2!}\mathtt{i}%
\lambda ^{2}\left[ Q_{i}A_{ij}P_{j},Q_{l}A_{lk}\right] +\cdots   \notag \\
&=&Q_{i}(e^{-\lambda A})_{ik}=(e^{-\lambda \tilde{A}})_{ki}Q_{i},  \label{9}
\\
S_{n}^{-1}P_{k}S_{n} &=&P_{k}+\lambda A_{ki}P_{i}+\frac{1}{2!}\mathtt{i}%
\lambda ^{2}\left[ A_{ki}P_{j},Q_{l}A_{lm}P_{m}\right] +\cdots   \notag \\
&=&(e^{\lambda A})_{ki}P_{i}.  \label{9a}
\end{eqnarray}%
From Eq.(\ref{9}) we see that when $S_{n}$ acts on the n-mode coordinate
eigenstate $\left \vert \vec{q}\right \rangle ,$ where $\widetilde{\vec{q}}%
=(q_{1},q_{2},\cdots ,q_{n})$, it squeezes $\left \vert \vec{q}\right \rangle $
in this way:
\begin{equation}
S_{n}\left \vert \vec{q}\right \rangle =\left \vert \Lambda \right \vert
^{1/2}\left \vert \Lambda \vec{q}\right \rangle ,\text{ }\Lambda =e^{-\lambda
\tilde{A}},\text{ }\left \vert \Lambda \right \vert \equiv \det \Lambda .
\label{10}
\end{equation}%
Thus $S_{n}$ has the representation on the Dirac's coordinate basis $%
\left \langle \vec{q}\right \vert $\cite{8b}
\begin{equation}
S_{n}=\int d^{n}qS_{n}\left \vert \vec{q}\right \rangle \left \langle \vec{q}%
\right \vert =\left \vert \Lambda \right \vert ^{1/2}\int d^{n}q\left \vert
\Lambda \vec{q}\right \rangle \left \langle \vec{q}\right \vert ,\text{ \  \ }%
S_{n}^{\dagger }=S_{n}^{-1},  \label{11}
\end{equation}%
since $\int d^{n}q\left \vert \vec{q}\right \rangle \left \langle \vec{q}%
\right \vert =1.$ Using the expression of $\left \vert \vec{q}\right \rangle $
in Fock space
\begin{eqnarray}
&\left \vert \vec{q}\right \rangle =\pi ^{-n/4}\colon \exp \left[ -\frac{1}{2}%
\widetilde{\vec{q}}\vec{q}+\sqrt{2}\widetilde{\vec{q}}a^{\dag }-\frac{1}{2}%
\tilde{a}^{\dag }a^{\dag }\right] \left \vert \mathbf{0}\right \rangle ,\text{
}&  \notag \\
&\tilde{a}^{\dag }=(a_{1}^{\dag },a_{2}^{\dag },\cdots ,a_{n}^{\dag })\text{,%
}&  \label{12}
\end{eqnarray}%
and the normally ordered form of n-mode vacuum projector $\left \vert \mathbf{%
0}\right \rangle \left \langle \mathbf{0}\right \vert =\colon \exp [-\tilde{a}%
^{\dag }a^{\dag }]\colon $, we can put $S_{n}$ into the normal ordering
form,
\begin{eqnarray}
S_{n} &=&\pi ^{-n/2}\left \vert \Lambda \right \vert ^{1/2}\int d^{n}q\colon
\exp [-\frac{1}{2}\widetilde{\vec{q}}(1+\widetilde{\Lambda }\Lambda )\vec{q}+%
\sqrt{2}\widetilde{\vec{q}}(\widetilde{\Lambda }a^{\dag }+a)  \notag \\
&&-\frac{1}{2}(\widetilde{a}a+\tilde{a}^{\dag }a^{\dag })-\tilde{a}^{\dag
}a]\colon .  \label{13}
\end{eqnarray}%
To perform the integration in Eq.(\ref{13}) by virtue of the IWOP technique,
using the mathematical formula
\begin{equation}
\int d^{n}x\exp [-\widetilde{x}Fx+\widetilde{x}v]=\pi ^{n/2}(\det
F)^{-1/2}\exp \left[ \frac{1}{4}\widetilde{v}F^{-1}v\right] ,  \label{14}
\end{equation}%
then we derive
\begin{eqnarray}
S_{n} &=&\left( \frac{\det \Lambda }{\det N}\right) ^{1/2}\exp \left[ \frac{1%
}{2}\tilde{a}^{\dag }\left( \Lambda N^{-1}\widetilde{\Lambda }-I\right)
a^{\dag }\right]   \notag \\
&&\times \colon \exp \left[ \tilde{a}^{\dag }\left( \Lambda N^{-1}-I\right) a%
\right] \colon \exp \left[ \frac{1}{2}\widetilde{a}\left( N^{-1}-I\right) a%
\right] ,  \label{15}
\end{eqnarray}%
where $N=(1+\widetilde{\Lambda }\Lambda )/2$. Eq.(\ref{15}) is just the
normal product form of $S_{n}.$

\section{Squeezing property of $S_{n}\left \vert \mathbf{0}\right \rangle $}

Operating $S_{n}$ on the n-mode vacuum state $\left \vert \mathbf{0}%
\right \rangle ,$ we obtain the squeezed vacuum state%
\begin{equation}
S_{n}\left \vert \mathbf{0}\right \rangle =\left( \frac{\det \Lambda }{\det N}%
\right) ^{1/2}\exp \left[ \frac{1}{2}\tilde{a}^{\dag }\left( \Lambda N^{-1}%
\widetilde{\Lambda }-I\right) a^{\dag }\right] \left \vert \mathbf{0}%
\right \rangle .  \label{16}
\end{equation}%
Now we evaluate the variances of the n-mode quadratures. The quadratures in
the n-mode case are defined as%
\begin{equation}
X_{1}=\frac{1}{\sqrt{2n}}\sum_{i=1}^{n}Q_{i},\text{ }X_{2}=\frac{1}{\sqrt{2n}%
}\sum_{i=1}^{n}P_{i},  \label{17}
\end{equation}%
obeying $[X_{1},X_{2}]=\frac{\mathtt{i}}{2}.$ Their variances are $\left(
\Delta X_{i}\right) ^{2}=\left \langle X_{i}^{2}\right \rangle -\left \langle
X_{i}\right \rangle ^{2}$, $i=1,2.$ Noting the expectation values of $X_{1}$
and $X_{2}$ in the state $S_{n}\left \vert \mathbf{0}\right \rangle $, $%
\left \langle X_{1}\right \rangle =\left \langle X_{2}\right \rangle =0$, then
using Eqs. (\ref{9}) and (\ref{9a}) we see that the variances are%
\begin{eqnarray}
\left( \triangle X_{1}\right) ^{2} &=&\left \langle \mathbf{0}\right \vert
S_{n}^{-1}X_{1}^{2}S_{n}\left \vert \mathbf{0}\right \rangle =\frac{1}{2n}%
\left \langle \mathbf{0}\right \vert
S_{n}^{-1}\sum_{i=1}^{n}Q_{i}\sum_{j=1}^{n}Q_{j}S_{n}\left \vert \mathbf{0}%
\right \rangle   \notag \\
&=&\frac{1}{2n}\left \langle \mathbf{0}\right \vert
\sum_{i=1}^{n}Q_{k}(e^{-\lambda A})_{ki}\sum_{j=1}^{n}(e^{-\lambda \tilde{A}%
})_{jl}Q_{l}\left \vert \mathbf{0}\right \rangle   \notag \\
&=&\frac{1}{2n}\underset{i,j}{\sum^{n}}(e^{-\lambda A})_{ki}(e^{-\lambda
\tilde{A}})_{jl}\left \langle \mathbf{0}\right \vert Q_{k}Q_{l}\left \vert
\mathbf{0}\right \rangle   \notag \\
&=&\frac{1}{4n}\underset{i,j}{\sum^{n}}(e^{-\lambda A})_{ki}(e^{-\lambda
\tilde{A}})_{jl}\left \langle \mathbf{0}\right \vert a_{k}a_{l}^{\dagger
}\left \vert \mathbf{0}\right \rangle   \notag \\
&=&\frac{1}{4n}\underset{i,j}{\sum^{n}}(e^{-\lambda A})_{ki}(e^{-\lambda
\tilde{A}})_{jl}\delta _{kl}=\frac{1}{4n}\underset{i,j}{\sum^{n}}(\widetilde{%
\Lambda }\Lambda )_{ij},  \label{18}
\end{eqnarray}%
similarly we have%
\begin{equation}
\left( \triangle X_{2}\right) ^{2}=\left \langle \mathbf{0}\right \vert
S_{n}^{-1}X_{2}^{2}S_{n}\left \vert \mathbf{0}\right \rangle =\frac{1}{4n}%
\underset{i,j}{\sum^{n}}\left[ (\widetilde{\Lambda }\Lambda )^{-1}\right]
_{ij}.  \label{19}
\end{equation}%
Eqs. (\ref{18}) -(\ref{19}) are the quadrature variance formula in the
transformed vacuum state acted by the operator $\exp [\mathtt{i}\lambda
Q_{i}A_{ij}P_{j}].$ By observing that $A$ in (\ref{8}) is a symmetric
matrix, we see%
\begin{equation}
\underset{i,j}{\sum^{n}}\left[ (A+\tilde{A})^{l}\right] _{i\text{ }%
j}=2^{2l}n,  \label{20}
\end{equation}%
then using $A\tilde{A}=\tilde{A}A,$ so $\widetilde{\Lambda }\Lambda
=e^{-\lambda (A+\tilde{A})}$, a symmetric matrix, we have
\begin{equation}
\underset{i,j=1}{\sum^{n}}(\widetilde{\Lambda }\Lambda )_{i\text{ }%
j}=\sum_{l=0}^{\infty }\frac{(-\lambda )^{l}}{l!}\underset{i,j}{\sum^{n}}%
\left[ (A+\tilde{A})^{l}\right] _{i\text{ }j}=n\sum_{l=0}^{\infty }\frac{%
(-\lambda )^{l}}{l!}2^{2l}=ne^{-4\lambda },  \label{21}
\end{equation}%
and%
\begin{equation}
\underset{i,j=1}{\sum^{n}}(\widetilde{\Lambda }\Lambda )_{i\text{ }%
j}^{-1}=ne^{4\lambda }.  \label{22}
\end{equation}%
It then follows%
\begin{eqnarray}
\left( \triangle X_{1}\right) ^{2} &=&\frac{1}{4n}\underset{i,j}{\sum^{n}}(%
\widetilde{\Lambda }\Lambda )_{ij}=\frac{e^{-4\lambda }}{4},  \label{23} \\
\left( \triangle X_{2}\right) ^{2} &=&\frac{1}{4n}\underset{i,j}{\sum^{n}}%
\left[ (\widetilde{\Lambda }\Lambda )^{-1}\right] _{ij}=\frac{e^{4\lambda }}{%
4}.  \label{23a}
\end{eqnarray}%
This leads to $(\triangle X_{1})(\triangle X_{2})=\frac{1}{4},$ which shows
that $S_{n}$ is a correct n-mode squeezing operator for the n-mode
quadratures in Eq.(\ref{17}). Furthermore, Eqs.(\ref{23}) and (\ref{23a})
clearly indicate that the squeezed vacuum state $S_{n}\left \vert \mathbf{0}%
\right \rangle $ may exhibit stronger squeezing ($e^{-4\lambda }$) in one
quadrature than that ($e^{-2\lambda }$) of the usual two-mode squeezed
vacuum state. This is a way of enhancing squeezing.

\section{The Wigner function of $S_{n}\left \vert \mathbf{0}\right \rangle $}

Wigner distribution functions \cite{10} of quantum states are widely studied
in quantum statistics and quantum optics. Now we derive the expression of
the Wigner function of $S_{n}\left \vert \mathbf{0}\right \rangle .$ Here we
take a new method to do it. Recalling that in Ref. \cite{11} we have
introduced the Weyl ordering form of single-mode Wigner operator $\Delta
_{1}\left( q_{1},p_{1}\right) $,%
\begin{equation}
\Delta _{1}\left( q_{1},p_{1}\right)
=\genfrac{}{}{0pt}{}{:}{:}\delta \left( q_{1}-Q_{1}\right) \delta
\left( p_{1}-P_{1}\right) \genfrac{}{}{0pt}{}{:}{:},  \label{24}
\end{equation}%
its normal ordering form is%
\begin{equation}
\Delta _{1}\left( q_{1},p_{1}\right) =\frac{1}{\pi }\colon \exp \left[
-\left( q_{1}-Q_{1}\right) ^{2}-\left( p_{1}-P_{1}\right) ^{2}\right] \colon
\label{25}
\end{equation}%
where the symbols $\colon \colon $ and
$\genfrac{}{}{0pt}{}{:}{:}\genfrac{}{}{0pt}{}{:}{:}$ denote the
normal ordering and the Weyl ordering, respectively. Note that the
order of Bose operators $a_{1}$ and $a_{1}^{\dagger }$ within a
normally ordered product and a Weyl ordered product can be permuted.
That is to say, even though $[a_{1},a_{1}^{\dagger }]=1$, we can
have $\colon a_{1}a_{1}^{\dagger
}\colon =\colon a_{1}^{\dagger }a_{1}\colon $ and$\genfrac{}{}{0pt}{}{:}{:}%
a_{1}a_{1}^{\dagger }\genfrac{}{}{0pt}{}{:}{:}=\genfrac{}{}{0pt}{}{:}{:}a_{1}^{\dagger }a_{1}\genfrac{}{}{0pt}{}{:}{%
:}.$ The Weyl ordering has a remarkable property, i.e., the
order-invariance
of Weyl ordered operators under similar transformations, which means%
\begin{equation}
U\genfrac{}{}{0pt}{}{:}{:}\left( \circ \circ \circ \right) \genfrac{}{}{0pt}{}{:}{:}U^{-1}=\genfrac{}{}{0pt}{}{:}{:}%
U\left( \circ \circ \circ \right) U^{-1}\genfrac{}{}{0pt}{}{:}{:},
\label{26}
\end{equation}%
as if the \textquotedblleft fence"
$\genfrac{}{}{0pt}{}{:}{:}\genfrac{}{}{0pt}{}{:}{:}$did not exist.

For n-mode case, the Weyl ordering form of the Wigner operator is
\begin{equation}
\Delta _{n}\left( \vec{q},\vec{p}\right) =\genfrac{}{}{0pt}{}{:}{:}\delta \left( \vec{q}-%
\vec{Q}\right) \delta \left( \vec{p}-\vec{P}\right)
\genfrac{}{}{0pt}{}{:}{:}, \label{27}
\end{equation}%
where $\widetilde{\vec{Q}}=(Q_{1},Q_{2},\cdots ,Q_{n})$ and $\widetilde{\vec{%
P}}=(P_{1},P_{2},\cdots ,P_{n})$. Then according to the Weyl ordering
invariance under similar transformations and Eqs.(\ref{9}) and (\ref{9a}) we
have%
\begin{eqnarray}
S_{n}^{-1}\Delta _{n}\left( \vec{q},\vec{p}\right) S_{n} &=&S_{n}^{-1}\genfrac{}{}{0pt}{}{%
:}{:}\delta \left( \vec{q}-\vec{Q}\right) \delta \left( \vec{p}-\vec{P}%
\right) \genfrac{}{}{0pt}{}{:}{:}S_{n}  \notag \\
&=&\genfrac{}{}{0pt}{}{:}{:}\delta \left( q_{k}-(e^{-\lambda
\tilde{A}})_{ki}Q_{i}\right)
\delta \left( p_{k}-(e^{\lambda A})_{ki}P_{i}\right) \genfrac{}{}{0pt}{}{:}{:}  \notag \\
&=&\genfrac{}{}{0pt}{}{:}{:}\delta \left( e^{\lambda
\tilde{A}}\vec{q}-\vec{Q}\right)
\delta \left( e^{-\lambda A}\vec{p}-\vec{P}\right) \genfrac{}{}{0pt}{}{:}{:}  \notag \\
&=&\Delta \left( e^{\lambda \tilde{A}}\vec{q},e^{-\lambda
A}\vec{p}\right) . \label{28}
\end{eqnarray}%
Thus using Eqs.(\ref{24}) and (\ref{28}) the Wigner function of $%
S_{n}\left
\vert \mathbf{0}\right \rangle $ is
\begin{eqnarray}
&&\left \langle \mathbf{0}\right \vert S_{n}^{-1}\Delta _{n}\left( \vec{q},%
\vec{p}\right) S_{n}\left \vert \mathbf{0}\right \rangle  \notag \\
&=&\frac{1}{\pi ^{n}}\left \langle \mathbf{0}\right \vert \colon \exp
[-(e^{\lambda \tilde{A}}\vec{q}-\vec{Q})^{2}-(e^{-\lambda A}\vec{p}-\vec{P}%
)^{2}]\colon \left \vert \mathbf{0}\right \rangle  \notag \\
&=&\frac{1}{\pi ^{n}}\exp [-(e^{\lambda \tilde{A}}\vec{q})^{2}-\left(
e^{-\lambda A}\vec{p}\right) ^{2}]  \notag \\
&=&\frac{1}{\pi ^{n}}\exp \left[ -\widetilde{\vec{q}}e^{\lambda A}e^{\lambda
\tilde{A}}\vec{q}-\widetilde{\vec{p}}e^{-\lambda \tilde{A}}e^{-\lambda A}%
\vec{p}\right]  \notag \\
&=&\frac{1}{\pi ^{n}}\exp \left[ -\widetilde{\vec{q}}\left( \Lambda
\widetilde{\Lambda }\right) ^{-1}\vec{q}-\widetilde{\vec{p}}\Lambda
\widetilde{\Lambda }\vec{p}\right] ,  \label{29}
\end{eqnarray}%
From Eq.(\ref{29}) we see that once the explicit expression of $\Lambda
\tilde{\Lambda}=\exp [-\lambda (A+\tilde{A})]$ is deduced, the Wigner
function of $S_{n}\left \vert \mathbf{0}\right \rangle $ can be calculated.

\section{Some examples of calculating the Wigner function}

For $n=2,$ form Eq.(\ref{6}) we have $S_{2}^{\prime }=\exp \left[ \mathtt{i}%
2\lambda \left( Q_{1}P_{2}+Q_{2}P_{1}\right) \right] $ which exhibits
clearly the stronger squeezing than the usual two-mode squeezing operator $%
S_{2}^{\prime }.$ For $n=3,$ the three-mode operator \cite{9} $S_{3}$, from
Eq.(\ref{8}) we see that the matrix $A$ is $\left(
\begin{array}{ccc}
0 & 1 & 1 \\
1 & 0 & 1 \\
1 & 1 & 0%
\end{array}%
\right) ,$ thus we have%
\begin{equation}
\Lambda \tilde{\Lambda}=\allowbreak \left(
\begin{array}{ccc}
u & v & \allowbreak v \\
\allowbreak v & u & \allowbreak v \\
v & v & u%
\end{array}%
\right) ,\text{ }u=\frac{2}{3}e^{2\lambda }+\frac{1}{3e^{4\lambda }},\text{ }%
v=\frac{1}{3e^{4\lambda }}-\frac{1}{3}e^{2\lambda },  \label{30}
\end{equation}%
and$\  \left( \Lambda \tilde{\Lambda}\right) ^{-1}$ is obtained by replacing $%
\lambda $ with $-\lambda $ in $\Lambda \tilde{\Lambda}.$ Thus the squeezing
state $S_{3}\left \vert 000\right \rangle $ is
\begin{equation}
S_{3}\left \vert 000\right \rangle =A_{3}\exp \left[ \frac{1}{6}%
A_{1}\sum_{i=1}^{3}a_{i}^{\dagger 2}-\frac{2}{3}A_{2}\sum_{i<j}^{3}a_{i}^{%
\dagger }a_{j}^{\dagger }\right] \left \vert 000\right \rangle ,
\end{equation}%
where
\begin{equation}
A_{1}=\left( 1-\text{sech}2\lambda \right) \tanh \lambda ,\text{ }A_{2}=%
\frac{\sinh 3\lambda }{2\cosh \lambda \cosh 2\lambda },A_{3}=\text{sech}%
\lambda \cosh ^{-1/2}2\lambda .
\end{equation}%
In particular, for the case of the infinite squeezing $\lambda \rightarrow
\infty $, Eq.(\ref{s3}) reduces to
\begin{equation}
S_{3}\left \vert 000\right \rangle \sim \exp \left \{ \frac{1}{6}\left[
\sum_{i=1}^{3}a_{i}^{\dagger 2}-4\sum_{i<j}^{3}a_{i}^{\dagger
}a_{j}^{\dagger }\right] \right \} \left \vert 000\right \rangle \equiv
\left \vert \  \right \rangle _{s_{3}},  \label{s3}
\end{equation}%
which is just the common eigenvector of the three compatible Jacobian
operators in three-body case with zero eigenvalues \cite{12}, i.e.,%
\begin{align}
\left( P_{1}+P_{2}+P_{3}\right) \left \vert \  \right \rangle _{s_{3}}& =0,%
\text{ }\left( Q_{3}-Q_{2}\right) \left \vert \  \right \rangle _{s_{3}}=0,
\notag \\
\text{ }\left( \frac{\mu _{3}Q_{3}+\mu _{2}Q_{2}}{\mu _{3}+\mu _{2}}%
-Q_{1}\right) \left \vert \  \right \rangle _{s_{3}}& =0,\text{ }\left( \mu
_{i}=\frac{m_{i}}{m_{1}+m_{2}+m_{3}}\right) ,
\end{align}%
as common eigenvector
\begin{equation}
\left[ P_{1}+P_{2}+P_{3},Q_{3}-Q_{2}\right] =0,\left[ \frac{\mu
_{3}Q_{3}+\mu _{2}Q_{2}}{\mu _{3}+\mu _{2}}-Q_{1},P_{1}+P_{2}+P_{3}\right]
=0.
\end{equation}%
Since the common eigenvector of three compatible Jacobian operators is an
entangled state, the state $\left \vert \  \right \rangle _{s_{3}}$ is also
an entangled state.$\ $

By using Eq.(\ref{29}) the Wigner function is%
\begin{eqnarray}
&&\left \langle \mathbf{0}\right \vert S_{3}^{-1}\Delta _{3}\left( \vec{q},%
\vec{p}\right) S_{3}\left \vert \mathbf{0}\right \rangle  \notag \\
&=&\frac{1}{\pi ^{3}}\exp \left[ -\frac{2}{3}\left( \cosh 4\lambda +2\cosh
2\lambda \right) \sum_{i=1}^{3}\left \vert \alpha _{i}\right \vert ^{2}%
\right]  \notag \\
&&\times \exp \left \{ -\frac{1}{3}\allowbreak \left( \sinh 4\lambda -2\sinh
2\lambda \right) \sum_{i=1}^{3}\alpha _{i}^{2}\right.  \notag \\
&&-\left. \frac{2}{3}\sum_{j>i=1}^{3}\left[ \left( \cosh 4\lambda -\cosh
2\lambda \right) \alpha _{i}\alpha _{j}^{\ast }+\left( \allowbreak \sinh
2\lambda +\sinh 4\lambda \right) \alpha _{i}\alpha _{j}\right] +c.c.\right
\} .  \label{31}
\end{eqnarray}

For $n=4$ case, the four-mode operator $S_{4}$ is

\begin{equation}
S_{4}=\exp \{ \mathtt{i}\lambda \left[ \left( Q_{1}+Q_{3}\right) \left(
P_{4}+P_{2}\right) +\left( Q_{2}+Q_{4}\right) \left( P_{1}+P_{3}\right) %
\right] \}  \label{5c}
\end{equation}%
the matrix $A=\left(
\begin{array}{cccc}
0 & 1 & 0 & 1 \\
1 & 0 & 1 & 0 \\
0 & 1 & 0 & 1 \\
1 & 0 & 1 & 0%
\end{array}%
\right) $ , thus we have
\begin{equation}
\Lambda \tilde{\Lambda}=\allowbreak \left(
\begin{array}{cccc}
r & t & s & t \\
t & r & t & s \\
s & t & r & t \\
t & s & t & r%
\end{array}%
\right) ,  \label{32}
\end{equation}%
where $r=\cosh ^{2}2\lambda ,s=\sinh ^{2}2\lambda ,t=-\sinh 2\lambda \cosh
2\lambda .$ Then substituting Eq.(\ref{32}) into Eq.(\ref{29}) we obtain%
\begin{equation}
\left \langle \mathbf{0}\right \vert S_{4}^{-1}\Delta _{4}\left( \vec{q},%
\vec{p}\right) S_{4}\left \vert \mathbf{0}\right \rangle =\frac{1}{\pi ^{4}}%
\exp \left \{ -2\cosh ^{2}2\lambda \left[ \sum_{i=1}^{4}\left \vert \alpha
_{i}\right \vert ^{2}+\left( M+M^{\ast }\right) \tanh ^{2}2\lambda +\left(
R^{\ast }+R\right) \allowbreak \tanh 2\lambda \right] \right \} ,  \label{33}
\end{equation}%
where $M=\alpha _{1}\alpha _{3}^{\ast }+\alpha _{2}\alpha _{4}^{\ast },$ $%
R=\alpha _{1}\alpha _{2}+\alpha _{1}\alpha _{4}+\alpha _{2}\alpha
_{3}+\alpha _{3}\alpha _{4}.$ This form differs evidently from the Wigner
function of the direct-product of usual two two-mode squeezed states' Wigner
functions. In addition, using Eq. (\ref{32}) we can check Eqs.(\ref{23}) and
(\ref{23a}). Further, using Eq.(\ref{32}) we have

\begin{equation}
N^{-1}=\frac{1}{2}\allowbreak \left(
\begin{array}{cccc}
2 & \tanh 2\lambda & 0 & \tanh 2\lambda \\
\tanh 2\lambda & 2 & \tanh 2\lambda & 0 \\
0 & \tanh 2\lambda & 2 & \tanh 2\lambda \\
\tanh 2\lambda & 0 & \tanh 2\lambda & 2%
\end{array}%
\right) ,\text{ }\det N=\cosh ^{2}2\lambda .  \label{34}
\end{equation}%
Then substituting Eqs.(\ref{34}) into Eq.(\ref{15}) yields the four-mode
squeezed state \cite{9,13},%
\begin{equation}
S_{4}\left \vert 0000\right \rangle =\text{sech}2\lambda \exp \left[ -\frac{1%
}{2}\left( a_{1}^{\dag }+a_{3}^{\dag }\right) \left( a_{2}^{\dag
}+a_{4}^{\dag }\right) \tanh 2\lambda \right] \left \vert 0000\right \rangle
,  \label{35}
\end{equation}%
from which one can see that the four-mode squeezed state is not the same as
the direct product of two two-mode squeezed states in Eq.(\ref{1}).

In summary, by virtue of Dirac's coordinate representation and the IWOP
technique: we have shown that an n-mode squeezing operator $S_{n}\equiv \exp
[i\lambda \sum_{i=1}^{n}(Q_{i}P_{i+1}+Q_{i+1}P_{i}))],$ $(Q_{n+1}=Q_{1},\
P_{n+1}=P_{1}),$\ is an n-mode squeezing operator which enhances the
stronger squeezing for the n-mode quadratures \cite{14}. $S_{n}$'s normally
ordered expansion and new n-mode squeezed vacuum states are obtained.

\textbf{ACKNOWLEDGEMENT} Work supported by the National Natural Science
Foundation of China under grants 10775097 and 10874174.

\end{document}